\documentclass[10pt,english]{article}
\usepackage{mathptmx}
\usepackage[T1]{fontenc}
\usepackage[latin9]{inputenc}
\usepackage[letterpaper]{geometry}
\usepackage{color}
\usepackage{amsmath}
\usepackage{amssymb}
\usepackage{graphicx}
\usepackage{esint}
\usepackage[authoryear]{natbib}



\usepackage{babel}
\begin{document}

\title{A full, self-consistent, treatment of thermal wind balance on oblate
fluid planets}

\author{Eli Galanti$^{1}$, Yohai Kaspi$^{1}$, and Eli Tziperman$^{2}$}

\date{\textit{\scriptsize{}$^{1}$ Department of Earth and Planetary Sciences,
Weizmann Institute of Science, Rehovot, Israel}\\
\textit{\scriptsize{}$^{2}$Department of Earth and Planetary Sciences,
Harvard University, Cambridge, MA, USA}}
\maketitle
\begin{center}
\textit{(Journal of Fluid Mechanics - under revision - July 2016) }
\par\end{center}

\section*{Abstract}

The nature of the flow below the cloud level on Jupiter and Saturn
is still unknown. Relating the flow on these planets to perturbations
in their density field is key to the analysis of the gravity measurements
expected from both the Juno (Jupiter) and Cassini (Saturn) spacecrafts
during 2016-18. Both missions will provide latitude-dependent gravity
fields, which in principle could be inverted to calculate the vertical
structure of the observed cloud-level zonal flow on these planets.
Theories to date connecting the gravity field and the flow structure
have been limited to potential theories under a barotropic assumption,
or estimates based on thermal wind balance that allow analyzing baroclinic
wind structures, but have made simplifying assumptions that neglected
several physical effects. These include the effects of the deviations
from spherical symmetry, the centrifugal force due to density perturbations,
and self-gravitational effects of the density perturbations. Recent
studies attempted to include some of these neglected terms, but lacked
an overall approach that is able to include all effects in a self-consistent
manner. The present study introduces such a self-consistent perturbation
approach to the thermal wind balance that incorporates all physical
effects, and applies it to several example wind structures, both barotropic
and baroclinic. The contribution of each term is analyzed, and the
results are compared in the barotropic limit to those of potential
theory. It is found that the dominant balance involves the original
simplified thermal wind approach. This balance produces a good order-of-magnitude
estimate of the gravitational moments, and is able, therefore, to
address the order one question of how deep the flows are given measurements
of gravitational moments. The additional terms are significantly smaller
yet can affect the gravitational moments to some degree. However,
none of these terms is dominant so any approximation attempting to
improve over the simplified thermal wind approach needs to include
all other terms. 

\section{Introduction}

The observed cloud-level flow on Jupiter and Saturn is dominated by
strong east-west (zonal) flows. The depth to which these flows extend
is unknown, and has been a topic of great debate over the past few
decades (see reviews by \citealp{Vasavada2005} and \citealp{Showman2016}).
One of the prime goals of the Juno mission to Jupiter and the Cassini
Grande Finale at Saturn is to estimate the depth of these flows via
precise gravitational measurements. If the flows are indeed deep,
and therefore perturb significant mass, then they can produce a gravity
signal that will be measurable \citep{Hubbard1999,Kaspi2010a}. Constraining
the depth of these flows will help explore the mechanisms driving
the jets \citep[e.g.,][]{Busse1976,Williams1978,Ingersoll1982,Cho1996,Showman2006,Scott2007,Kaspi2007,Lian2010,Liu2010,Heimpel2016},
and give better constraints on interior structure models \citep[e.g.,][]{Guillot2005,Militzer2008,Nettelmann2012a,Helled2013}.

Several studies over the past decades have examined the effects of
interior flow on the gravitational moments. The gravity moment spectrum
mostly results from the planet's oblate shape due to its rotation,
and from the corresponding interior density distribution. However,
density perturbations due to atmospheric dynamics and internal flows
can affect the measured gravity moments especially if the flows extend
deep enough into the planets. \citet{Hubbard1982} and \citet{Hubbard1999}
used potential theory to calculate the gravity moments due to internal
flows, by extending the observed cloud-level winds along cylinders
throughout the planet as suggested by \citet{Busse1976}. This approach
takes into account the oblateness of the planet, yet is only possible
for the barotropic case, meaning that the flow is constant along lines
parallel to the axis of rotation. This occurs if the baroclinicity
vector $\nabla\rho\times\nabla p$ vanishes (e.g., if density is a
function of pressure only) at small Rossby number and negligible dissipation.
More recently, Hubbard introduced more accurate calculations, for
the gravitational signature of the flows \citep{Hubbard2012,Kong2012,Hubbard2013},
using concentric Maclaurin spheroids (CMS), but these are also only
limited to the fully barotropic case.

A different approach, assuming the large scale flow is dominated by
the rotation of the planet, used thermal wind balance to calculate
the gravity moments due to the wind field \citep{Kaspi2010a,Kaspi2013a,Kaspi2013c,Liu2013,Liu2014}.
The thermal wind approach is not limited to the barotropic case (can
account for any wind field), and in the barotropic limit has been
shown to be equivalent to the potential theory and CMS methods \citep{Kaspi2016}.
In addition, this approach allows for the calculation of the odd gravitational
moments, which can emerge from north-south hemispherical asymmetries
in the wind structure \citep{Kaspi2013a}. However, this thermal wind
approach was originally limited to spherical symmetry, resulting in
an inability to calculate the effects of the planet oblateness on
the gravity signature of the winds. \citet{Cao2015} added the effects
of oblateness on the background state density and gravity and concluded
that it should be considered when estimating the effects of the winds
on the gravity moments using thermal-wind. Similarly, \citet{Zhang2015}
included another effect, of the gravity anomalies due to density perturbations
associated with the winds, and found an effect on the second gravity
moment $J_{2}$, terming their approach thermal-gravity wind (TGW)
method. However, while both of these recent studies found some effects
of the terms they added, their choice of added physics did not result
from a systematic and self-consistent approach.

The purpose of this study is to develop a full, self-consistent thermal
wind (FTW) perturbation approach for the treatment of the general
thermal wind balance on a fluid planet. This will allow to calculate
density anomalies and gravity moments due to prescribed winds, omitting
the traditional sphericity assumptions which have been adopted from
dynamics on terrestrial planets. Our approach includes the effect
of oblateness, as well as that of gravity anomalies due to the dynamical
density perturbations themselves. We show these two effects to be
but two of several different factors that should be considered in
a self-consistent calculation. Our approach is based on a systematic
perturbation expansion, which both allows us to consider all effects,
and also points the way to improving the estimated gravity moments
using a higher order perturbation that can be considered by future
studies.

By including all relevant terms in the general thermal wind balance,
we are able to evaluate the relative contribution of different terms.
We find that the simplified thermal wind (TW) approach captures most
of the relation between the wind shear and density gradients. The
term added in TGW is found to be one of several smaller terms that
all need to be added together for consistency in order to improve
the estimates of the simplified thermal wind balance. Furthermore,
previous applications of the thermal wind balance encountered an unknown
integration constant that was a function of radius only and could
not be solved for. We show that this integration constant may have
an effect, although small, on the gravity moments, and develop a method
for calculating it.

The following section describes the perturbation approach, the resulting
equations and how they are solved. Next, in section~\ref{sec:Results-for-wind-induced},
we first verify this approach by comparing it to the results of the
CMS method in the barotropic limit, and then compare our results to
the less complete approaches of simplified thermal wind and TGW. We
then also apply the self-consistent solution to a case with baroclinic
winds where CMS cannot be used. We discuss the results and conclude
in section~\ref{sec:Discussion-and-conclusion}.

\section{Methods: perturbation expansion of the momentum equations\label{sec:Perturbation-expansion}}

We begin by taking the standard form of the momentum equations on
a planet rotating at an angular velocity $\Omega$, 
\begin{eqnarray}
\frac{\partial\mathbf{u}}{\partial t}+\left(\mathbf{u\cdot\nabla}\right)\mathbf{u}+2\boldsymbol{\Omega}\times\mathbf{u}+\boldsymbol{\Omega}\times\boldsymbol{\Omega}\mathbf{\times r} & = & -\frac{1}{\rho}\nabla p+\nabla\Phi,\label{eq:momentum}
\end{eqnarray}
where $\mathbf{u}$ is the 3D wind vector, $\rho$ is density, $p$
is pressure, $\boldsymbol{\Omega}$ is the planetary rotation rate,
and $\Phi$ is the body force potential. The first term on the lhs
is the local acceleration of the flow, the second is the Eulerian
advection, the third is the Coriolis acceleration, and the fourth
is the centrifugal acceleration. On right hand side appear the pressure
gradient term and the body force (gravity in this case, so that $\nabla\Phi={\bf -g}$).
Note that by gravity we refer here to the force due to the Newtonian
potential, not to the modified gravity which is commonly used in geostrophic
studies and includes the centrifugal potential as well. Typical values
for a Jupiter-like planet are $U=O\left(100\right)$~m~s$^{-1}$,
$\Omega=O\left(10^{-4}\right)$~s$^{-1}$, $a=O\left(7\times10^{7}\right)$~m,
where $a$ is the planet radius. The resulting Rossby number ($Ro$)
is therefore much smaller than one ($Ro\equiv U/\Omega a\thickapprox10^{-2}$),
and as the first two terms in Eq.~\ref{eq:momentum} can be neglected
so that the resulting balance is,
\begin{eqnarray}
2\boldsymbol{\Omega}\mathbf{\times\left(\rho u\right)} & = & -\nabla p-\rho{\bf g}-\rho\boldsymbol{\Omega}\times\boldsymbol{\Omega}\mathbf{\times r}.\label{eq:momentum-1}
\end{eqnarray}
Next, we denote the static solution ($\mathbf{u}=0$) as $\rho_{0},p_{0},g_{0}$,
and the perturbations due to the non zero wind (dynamical part of
the solution) as $\rho',p',g'$, such that
\begin{eqnarray}
\rho & = & \rho_{0}(r,\theta)+\rho'(r,\theta),\nonumber \\
p & = & p_{0}(r,\theta)+p'(r,\theta),\nonumber \\
{\bf g} & = & {\bf g}_{0}(r,\theta)+{\bf g}'(r,\theta).\label{eq:mean-anomalous-states}
\end{eqnarray}
Note that both the static and the dynamic solutions are functions
of latitude and radius, and that the gravity is directly related to
the density via a relation shown below. 

The equation obtained by setting the small Rossby number to zero as
a first approximation is effectively static and does not include the
velocity field,
\begin{eqnarray}
0 & = & -\nabla p_{0}-\rho_{0}g_{0}-\rho_{0}\boldsymbol{\Omega}\times\boldsymbol{\Omega}\mathbf{\times r},\label{eq:momentum-static}
\end{eqnarray}
and the dynamical perturbations therefore satisfy,
\begin{eqnarray}
2\boldsymbol{\Omega}\mathbf{\times\left(\rho_{0}u\right)} & = & -\nabla p'-\rho_{0}{\bf g}'-\rho'{\bf g}_{0}-\rho'\boldsymbol{\Omega}\times\boldsymbol{\Omega}\mathbf{\times r}.\label{eq:momentum-dynamic}
\end{eqnarray}
The solution procedure outlined here involves first finding the static
solution and then solving Eq.~\ref{eq:momentum-dynamic} for the
dynamical perturbations to the density due to the effects of the prescribed
winds. Taking the curl of Eq.~\ref{eq:momentum-dynamic} yields a
single equation in the azimuthal direction
\begin{eqnarray}
-2\Omega r\partial_{z}\left(\rho_{0}u\right) & = & -rg_{0}^{(\theta)}\frac{\partial\rho'}{\partial r}-g_{0}^{(r)}\frac{\partial\rho'}{\partial\theta}+r\frac{\partial\rho_{0}}{\partial r}g'{}^{(\theta)}-g'^{(r)}\frac{\partial\rho_{0}}{\partial\theta}\Omega^{2}r\left[\frac{\partial\rho'}{\partial\theta}\cos^{2}\theta+\frac{\partial\rho'}{\partial r}r\cos\theta\sin\theta\right],\label{eq:vorticity equation for dynamic state}
\end{eqnarray}
where the notation $\partial_{z}=\cos\theta\frac{\partial}{\partial r}-\sin\theta\frac{\partial}{\partial\theta}$
denotes the derivative along the direction of the axis of rotation
($z$), and gravity is expressed as function of the density as,

\begin{equation}
\mathbf{g}(r,\theta)\mathbf{=}2\pi G\left[\frac{\partial}{\partial r},\frac{\partial}{r\partial\theta}\right]\iint\left\langle \frac{1}{|\mathbf{r-}\mathbf{r}'|}\right\rangle \mathbf{\rho}(r',\theta')r'^{2}\cos\theta'd\theta'dr'\label{eq:grav-1}
\end{equation}
where the gravity can be either $\mathbf{g_{0}}$ or $\mathbf{g'}$,
calculated from $\rho_{0}$ or $\rho'$, respectively, and $G=6.67\times10^{-11}$~m$^{3}$$\,$kg$^{-1}\,$$\,$s$^{-2}$
is the gravitational constant. Note that Eq.~\ref{eq:vorticity equation for dynamic state},
together with Eq.~\ref{eq:grav-1}, form an integro-differential
equation whose solution requires the calculation of integration constants
as is done below and demonstrated in a simpler context in Appendix
B.

\begin{figure}
\begin{centering}
\includegraphics[scale=0.4]{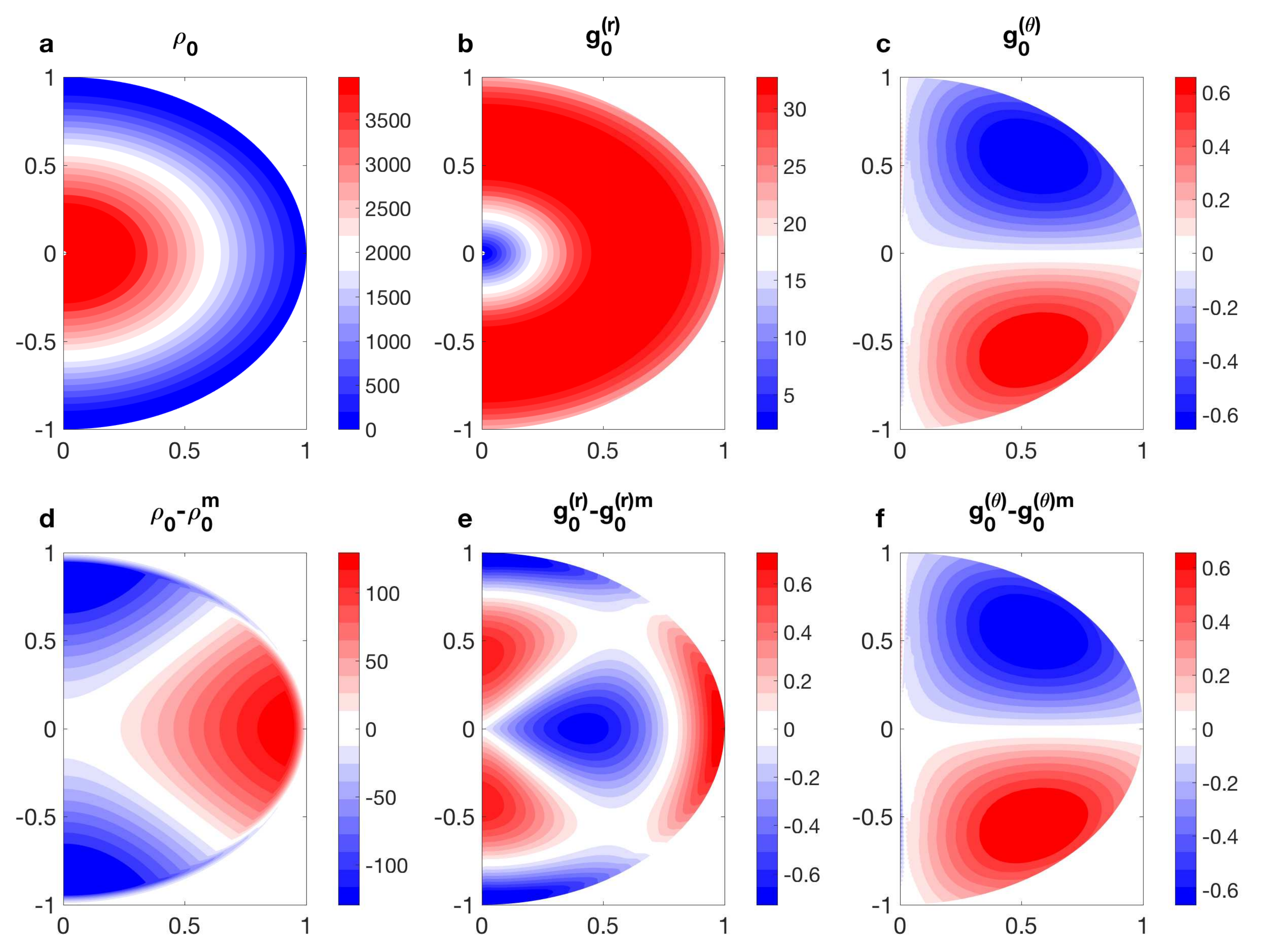}
\par\end{centering}
\caption{The static model solution, as function of radius and latitude, from
the CMS model for the density and gravity (panels a-c), and their
deviation from the latitudinal mean (panels d-f). \label{fig: background solution}}
\end{figure}

The above equations for the perturbation density are the first order
perturbation equations, which are solved in this study. It is important
to note, though, that because this is a self-consistent treatment
of the density perturbations, it also allows improving on the approximation
by proceeding to the next orders. As a demonstration, we write the
second order perturbation equations in Appendix~\ref{subsec:Appendix Higher-order-perturbation}.

\subsection{The background static solution}

The static density $\rho_{0}$ and gravity $\mathbf{g_{0}}$ are taken
from the solution of the CMS model \citep{Hubbard2012,Hubbard2013}.
The model is based on a numerical method for solving the equilibrium
shape of a rotating planet, for which an analytic solution exists
in the form of a Maclaurin spheroid. The continuously varying density
and pressure structures are represented by a discrete set of layers,
in which the density and pressure are constant. This onion-like structure
is then decomposed into a set of concentric Maclaurin spheroids, and
a solution is sought for by requiring that the sum of the gravitational
potential and the rotational potential are constant on the surface
of the planet \citep{Hubbard2012,Hubbard2013}. Solutions using this
method give similar results to other methods \citep{Wisdom2016,Kaspi2016}.
The static fields resulting from the CMS solution are shown in Fig.~\ref{fig: background solution}.
Both the density and gravity fields show a structure that is mostly
radial. While the density (Fig.~\ref{fig: background solution}a)
ranges from zero to about 4000~kg$\,$m$^{-3}$, its latitude-dependent
component (Fig.~\ref{fig: background solution}d) ranges only between
-150 and 100~kg$\,$m$^{-3}$. Similarly, the radial gravity is also
dominated by its radial component (Fig.~\ref{fig: background solution}b,e),
with the peak gravity being $\sim32$~m$\,$s$^{-2}$ at about $0.7a$.
The latitudinal component of the gravity is much weaker than the radial
component (Fig.~\ref{fig: background solution}c,f). Comparing panels
c and f, shows that the latitudinal mean of the latitudinal component
of the gravity is zero, and has a much smaller magnitude than the
radial component. Note that in Fig.~\ref{fig: background solution}c,f
a positive value means gravity pointing northward.

\subsection{Solving for the dynamic density perturbations}

We solve Eq.~\ref{eq:vorticity equation for dynamic state} by writing
the equations in matrix form \citep[e.g.,][]{Zhang2015}. The 2D problem
(radius and latitude) is discretized in both directions, where $N_{r}$
and $N_{\theta}$ are the number of grid points in radius and latitude,
respectively. $N=N_{r}\times N_{\theta}$ is the total number of grid
points. Equation~\ref{eq:vorticity equation for dynamic state} is
then written in matrix form,
\begin{eqnarray}
b & = & A\rho',\label{eq:numerical equation - rho'}
\end{eqnarray}
where $A$ is a $N\times N$ matrix with contributions from all terms
in the $rhs$ of the equation, $b$ is the known $lhs$ of the equation,
due to the prescribed wind field, written an a $N\times1$ vector,
and the unknown density perturbation $\rho'$ is written as a $N\times1$
vector. All partial derivatives are written in center finite difference
form, aside from near the boundaries where the derivatives are evaluated
between grid points and weighted together with the adjacent derivative.
Note that the gravity is fully calculated in the matrix (Eq.~\ref{eq:numerical equation - rho'}).
It is done explicitly for each grid point.

Solving Eq.~\ref{eq:numerical equation - rho'} involves the inversion
of the matrix $A$, which is not possible because the matrix is singular
and thus has a null space, which leads to a part of the solution that
cannot be determined from the equation. This can be better understood
in the simpler case where equation \ref{eq:vorticity equation for dynamic state}
is reduced to the thermal wind balance with a spherical base state,
involving only the \textit{lhs} and the second term on the \textit{rhs}.
In that case, the equation can be integrated in latitude, leaving
an unknown integration constant that is a function of radius alone
\citep[e.g., ][]{Kaspi2010a}. In the more general case solved here,
while the base state is a function of both radius and latitude, the
null space still has a dimension equal to the number of radial grid
points $N_{r}$, similar to the unknown function of radius only, but
which has latitudinal dependence as well. We show in Appendix~\ref{subsec:Appendix integration-constant}
how the integration constant is calculated analytically in the simpler
case, and discuss below (section~\ref{subsec:Calculating-drho})
the solution in the more general case. In order to solve for the non-null
part of the solution, we use singular value decomposition as follows.
Let
\[
A=U\Sigma V^{T},
\]
where $U$ and $V$ are unitary matrixes, and $\Sigma$ is a rectangular
diagonal matrix, then the pseudo inverse of $A$ is given by
\[
A^{\dagger}=V\Sigma^{\dagger}U^{T},
\]
where 
\begin{eqnarray*}
\Sigma & = & diag(\sigma_{1},\ldots,\sigma_{N-N_{r}},0,\ldots,0),\\
\Sigma^{\dagger} & = & diag(\sigma_{1}^{-1},\ldots,\sigma_{N-N_{r}}^{-1},0,\ldots,0)\text{.}
\end{eqnarray*}
The solution is now written as $\rho'=\hat{\rho}'+\delta\rho'$, such
that $\hat{\rho}'$ is the part obtained from the pseudo inverse,
\[
\hat{\rho}'=A^{\dagger}b,
\]
and the additional component $\delta\rho'$ is in the null space of
$A$, so that $A\rho'=A(\hat{\rho}'+\delta\rho')=A\hat{\rho}'=b$.
That is, denoting the eigenvectors of $A$ by $\mathbf{e}_{i}$, such
that $A\mathbf{e}_{i}=\lambda_{i}\mathbf{e}_{i}$, then the null space
of the solution corresponds to any linear combination of eigenvectors
corresponding to the zero eigenvalues, which may be added to the solution
$\hat{\rho'}$ while still satisfying $A\rho'=b$. We next discuss
the calculation of the null space and its contribution to the solution
for $\rho'$. 

\subsection{Calculating the null-space solution\label{subsec:Calculating-drho}}

\begin{figure}
\begin{centering}
\includegraphics[scale=0.45]{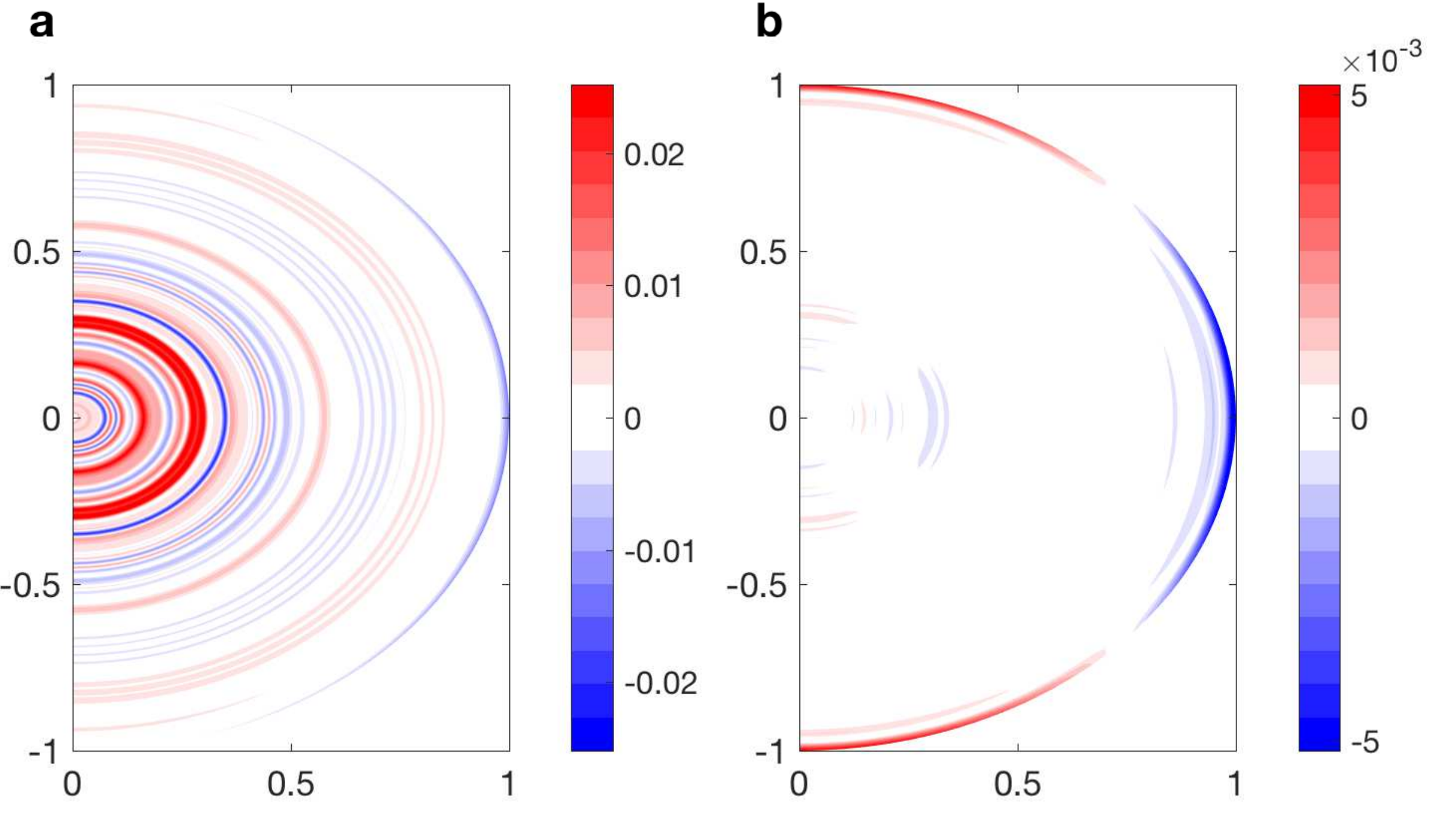}
\par\end{centering}
\caption{An example of a null mode $V$. (a) the full null model, (b) the latitudinal
dependent part of the null mode ($V_{a}=V-\overline{V})$. Note that
in most of the domain, the structure of the mode depends on radius
only, aside from the region close to the surface where a large scale
latitude dependent structure exists, but whose value is around half
the magnitude of the full structure. \label{fig: null modes example}}
\end{figure}

As mentioned above, the number of null eigenvalues is always found
to be $N_{n}=N_{r}$, i.e., equal to the number of grid points in
the radial direction, hinting to the possibility that the modes are
predominantly radially dependent. Indeed, looking at the zero eigenvectors
does show this characteristic (see example in Fig.~\ref{fig: null modes example}a).
Nevertheless, since the null eigenvectors do depend on latitude (Fig.~\ref{fig: null modes example}b),
and this dependence is concentrated close to the planet upper levels,
it could have a substantial effect on the gravity moments. The contribution
of the null space to the density may be written as,

\begin{eqnarray}
\delta\rho' & = & \sum_{i=N-N_{r}+1}^{N}c_{i}\mathbf{e}_{i}=E\mathbf{c},\label{eq:null-part-of-solution}
\end{eqnarray}
where $E$ is a $N\times N_{n}$ matrix whose columns are the $N_{n}$
null eigenvectors, and $\mathbf{c}$ is a $N_{n}\times1$ vector of
the unknown amplitudes of the null eigenvectors. In order to calculate
the vector $\mathbf{c}$, we must introduce additional physics in
the form of the polytropic relation between the pressure and the density
(a demonstration of a full analytical solution using the polytropic
equation is shown in Appendix~\ref{subsec:Appendix integration-constant}
for a simpler case). An additional constraint will be the conservation
of the total mass of the planet. Note that the polytropic relation
is only used for the solution of the null space, but the rest of the
solution is independent of it.

The polytropic equation, $p=K\rho^{2}$, linearized around the static
solution, gives 
\begin{equation}
p'=Kn\rho_{0}\rho',\label{eq:poly1}
\end{equation}
which may be used to define $\hat{p}'$ as
\begin{equation}
\hat{p}'\equiv Kn\rho_{0}\hat{\rho}'.\label{eq:poly2}
\end{equation}
Previously we took the curl of the momentum equation (Eq.~\ref{eq:momentum-dynamic}),
to solve for $\hat{\rho}'(r,\theta)$, and it satisfies the original
equation up to a gradient of some scalar function $\psi$. We therefore
may replace the pressure $\hat{p}'$ with $\hat{p}'+\psi$. Using
this augmented function, the perturbation momentum equation may be
written as, 
\begin{equation}
2\boldsymbol{\Omega}\mathbf{\times u}\rho_{0}=-\nabla(\hat{p}'+\psi)-\rho_{0}\mathbf{\hat{g}}'-\hat{\rho}'\mathbf{g}_{0}-\hat{\rho}'\boldsymbol{\Omega}\times\boldsymbol{\Omega}\mathbf{\times r}.\label{eq:poly_mom1}
\end{equation}
Next, taking the difference between the momentum equation for $\rho'$
and for $\hat{\rho}'$, we find,
\begin{equation}
0=-\nabla(p'-\hat{p}'-\psi)-\rho_{0}\mathbf{\delta g}'-\delta\rho'\mathbf{g}_{0}-\delta\rho'\boldsymbol{\Omega}\times\boldsymbol{\Omega}\mathbf{\times r}.\label{eq:poly_mom2}
\end{equation}
The function $\psi$ appears as a correction to the perturbation pressure,
and we therefore define the perturbation pressure to be, 
\[
\delta p'\equiv p'-\hat{p}'-\psi.
\]
Taking the difference between the polytropic equation for $\rho$
and $\hat{\rho}'$, we have
\[
\delta p'=Kn\rho_{0}\delta\rho',
\]
which leads to an equation for the unknown perturbation density,
\begin{equation}
0=-\nabla(Kn\rho_{0}\delta\rho')-\rho_{0}\mathbf{\delta g}'-\delta\rho'\mathbf{g}_{0}-\delta\rho'\boldsymbol{\Omega}\times\boldsymbol{\Omega}\mathbf{\times r},\label{eq:poly_mom3}
\end{equation}
 and explicitly, 
\begin{eqnarray*}
F_{r}:\,\,\,\,\,\,0 & = & -2K\frac{\partial}{\partial r}(\rho_{0}\delta\rho')-\rho_{0}\mathbf{\delta}g'^{(r)}-\delta\rho'g_{0}^{(r)}+\delta\rho'\Omega^{2}r\cos^{2}\theta,\\
F_{\theta}:\,\,\,\,\,\,0 & = & -2K\frac{\partial}{r\partial\theta}(\rho_{0}\delta\rho')-\rho_{0}\mathbf{\delta}g'^{(\theta)}-\delta\rho'g_{0}^{(\theta)}-\delta\rho'\Omega^{2}r\cos\theta\sin\theta,
\end{eqnarray*}
where $F_{r}$ and $F_{\theta}$ are the equations in the radial and
latitudinal directions, respectively. This represents the radial and
latitudinal components of a vector equation, and we next take the
divergence to get a single equation,

\begin{eqnarray}
0 & = & \frac{1}{r^{2}}\frac{\partial}{\partial r}(r^{2}F_{r})+\frac{1}{r\sin\theta}\frac{\partial}{\partial\theta}(\sin\theta F_{\theta}),\label{eq:delta rho function}
\end{eqnarray}
where everything but $\delta\rho'$ is known. Numerically, this equation
can be written as a set of linear equations, 

\begin{eqnarray*}
B\delta\rho' & = & 0,
\end{eqnarray*}
where $B$ is a $N\times N$ matrix, and the \textit{rhs} is a $N\times1$
vector with zeros in all entries. An additional constraint is that
the total mass of the planet must not change due to the existence
of the wind, so that $\int\rho'd^{3}\mathbf{r}=0$ and therefore,
\begin{equation}
\int\delta\rho'd^{3}\mathbf{r}=-\int\hat{\rho}'d^{3}\mathbf{r}=-\delta M.\label{eq:mass conservation condition}
\end{equation}
The mass $\delta M$, due to the non-null space solution $\hat{\rho}'$,
is known from the solution to Eq.~\ref{eq:vorticity equation for dynamic state}.
Adding this constraint to Eq.~\ref{eq:delta rho function} results
in augmenting the matrix $B$ with an additional row, to form a matrix
$\tilde{B}$ whose size is now $(N+1)\times N$ and the rhs, now denoted
$\mathbf{m}$, is now a vector of length $N+1$ with a nonzero value
only in the last entry, $m_{N+1}=-\delta M$. Using the definition
for the null space part of the solution (\ref{eq:null-part-of-solution})
we get,

\begin{eqnarray*}
\tilde{B}E\mathbf{c} & = & \mathbf{m}.
\end{eqnarray*}
This is a formally overdetermined problem, which is solved for $\mathbf{c}$
using least squares \citep{Strang2006}, allowing us to then calculate
$\delta\rho'$. Note that even though $\tilde{B},E$ and $\mathbf{m}$
are all complex, $\delta\rho'$ is found to be real, as expected.

The use of a polytropic relation between the pressure and density
implies that the baroclinic vector $\nabla p\times\nabla\rho$ vanishes,
and therefore that the velocity field is necessarily barotropic at
small Rossby numbers. This is in line with most of the cases discussed
here that are indeed barotropic, aside from the last case that is
baroclinic and is analyzed in section 3.3. Note that a different pressure-density
relation is also possible. Nonetheless, in all cases discussed here
the contribution of the null space solution to the overall solution
is negligible. Our purpose here is to show how adding information
regarding the equation of state (in this case $p(\rho)$) can be used
to calculate the unknown integration constant arising in the thermal
wind formulation \citep[e.g., ][]{Kaspi2010a,Zhang2015}. However,
in future application, one would need to use a more realistic equation
of state that allows for determining the null space for a baroclinic
wind field as well.

\subsection{Prescribed winds\label{subsec:Prescribed-winds}}

The wind profile used in this study is based on the measured cloud-tracking
wind during the Cassini flyby \citep{Porco2003}. Since we compare
our results to the CMS model solution as a reference for the full
oblate solution, and the CMS wind profile must be truncated for numerical
convergence (see \citet{Kaspi2016} for details), we use a 24th degree
expansion of its differential potential. \citet{Kaspi2016} shows
a comparison between the resulting gravity moments using the truncated
and untruncated wind profiles. The choice of the specific wind profile
does not affect the results. In order to do a proper comparison to
the CMS model, which is limited to barotropic winds, the wind profile
is extended along cylinders parallel to the direction of the axis
of rotation. For the baroclinic case discussed in (section~\ref{subsec:analysis-baroclinic-winds}),
the wind profile is extended toward the interior using an exponential
decay function (e.g.,~as in \citealp{Galanti2016}) with a decay
scale height of $1000$~km.

\subsection{Calculating gravity moments}

In all cases discussed below, in addition to examining the solution
for the density perturbations, we calculate the resulting gravitational
moments given by
\begin{equation}
\Delta J_{n}=-\frac{2\pi}{Ma^{n}}\intop_{0}^{a}r'^{n+2}dr'\intop_{-1}^{1}P_{n}\left(\mu'\right)\rho'\left(r',\mu'\right)d\mu',\label{eq: dynamical zonal harmonics}
\end{equation}
where $M$ is the mass of the planet, $P_{n}$ are the Legendre polynomials,
and $\mu=\cos(\theta)$. Note that any part of $\rho'$ that is a
function of radius only does not contribute to the gravity moments.
For instance, using the latitudinal average of the density, $\rho'_{m}(r)=\overline{\rho'(r,\theta)}$
in Eq.~\ref{eq: dynamical zonal harmonics} will give
\begin{equation}
\Delta J_{n}=-\frac{2\pi}{Ma^{n}}\intop_{0}^{a}r'^{n+2}\rho'_{m}\left(r'\right)dr'\intop_{-1}^{1}P_{n}\left(\mu'\right)d\mu'=0,\label{eq: dynamical zonal harmonics-constant of integration}
\end{equation}
which vanishes due to the Legendre polynomials having a zero latitudinal
mean for any value of $n$. Therefore, any solution for $\rho'$ needs
to be examined with respect to its latitudinal dependent part.

\section{Results for wind-induced density and gravity moments\label{sec:Results-for-wind-induced}}

We now consider the solution for the density field and gravitational
moments in several cases. First, we examine the case of barotropic
winds where the results of the perturbation approach can be compared
to the concentric Maclaurin spheroids (CMS) solution (section~\ref{subsec:Verification-of-perturbation}).
Second, we compare our approach to earlier methods and approximations
(Section~\ref{subsec:Analysis-of-solution}). Finally, we analyze
an example of the more general case of baroclinic winds, where a CMS
solution is not possible (section~\ref{subsec:analysis-baroclinic-winds}). 

\subsection{Verification of perturbation method via a comparison to CMS\label{subsec:Verification-of-perturbation}}

\begin{figure}
\begin{centering}
\includegraphics[scale=0.43]{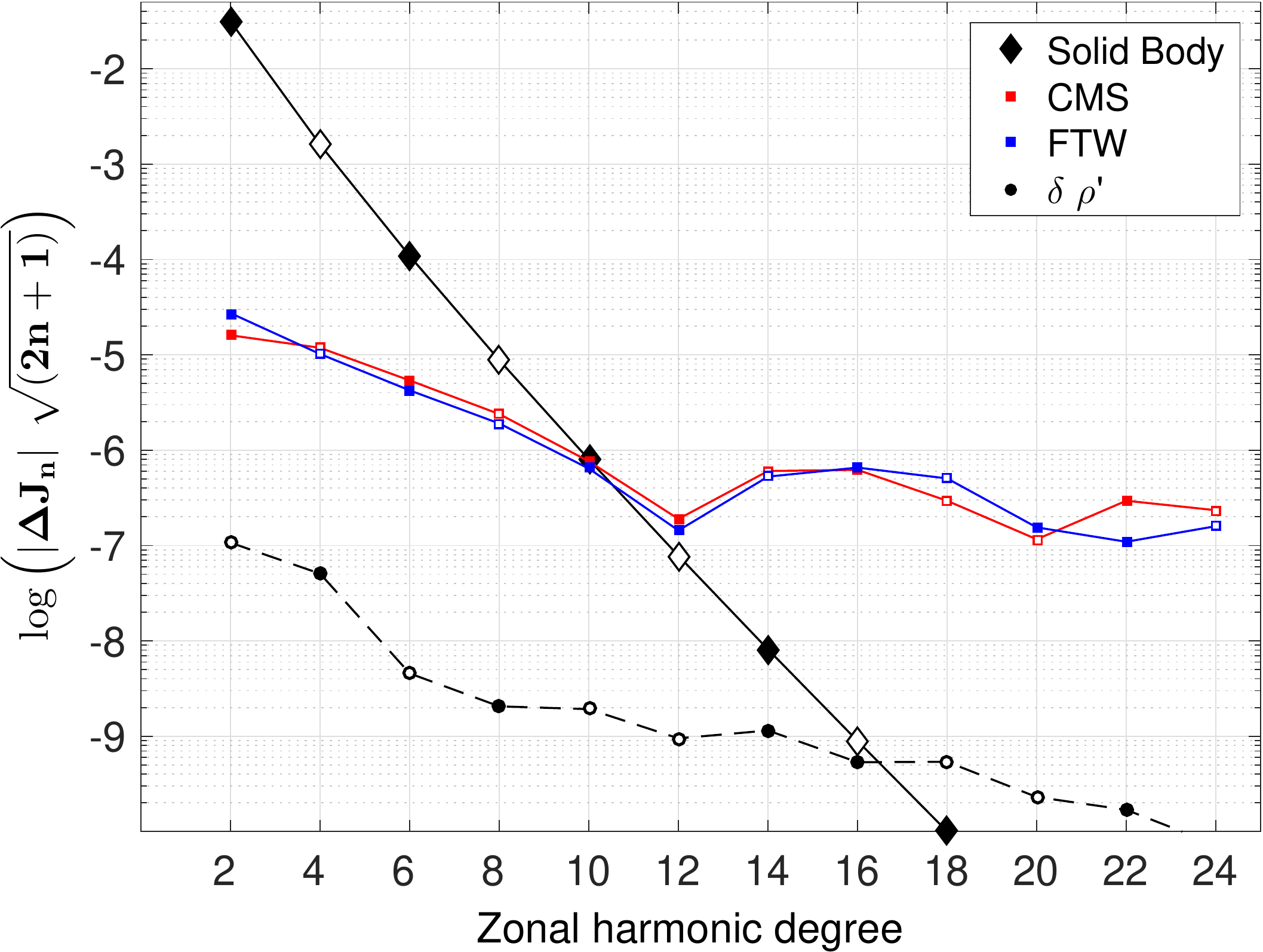}
\par\end{centering}
\caption{The wind induced gravitational moments Solution of Eq.~\ref{eq: dynamical zonal harmonics}
when all terms are included (blue), compared to the CMS solution (red).
Also shown are the contribution from $\delta\rho'$ (black-dashed
line), and the solid body induced gravitational moments (black). \label{fig: moments - full PA}}
\end{figure}

Solving numerically Eq.~\ref{eq:vorticity equation for dynamic state},
with all six terms on the \textit{rhs} included and adding the null
space solution, we obtain the anomalous density field $\rho'=\hat{\rho}+\delta\rho'$
from which we calculate the gravitational moments shown in Fig.~\ref{fig: moments - full PA}
(blue line), together with the reference CMS solution (red line).
The perturbation analysis captures most of the signal of the moments.
The dashed line shows the contribution of the null mode solution $\delta\rho'$
that is much smaller than the total solution.

\begin{figure*}
\begin{centering}
\includegraphics[scale=0.5]{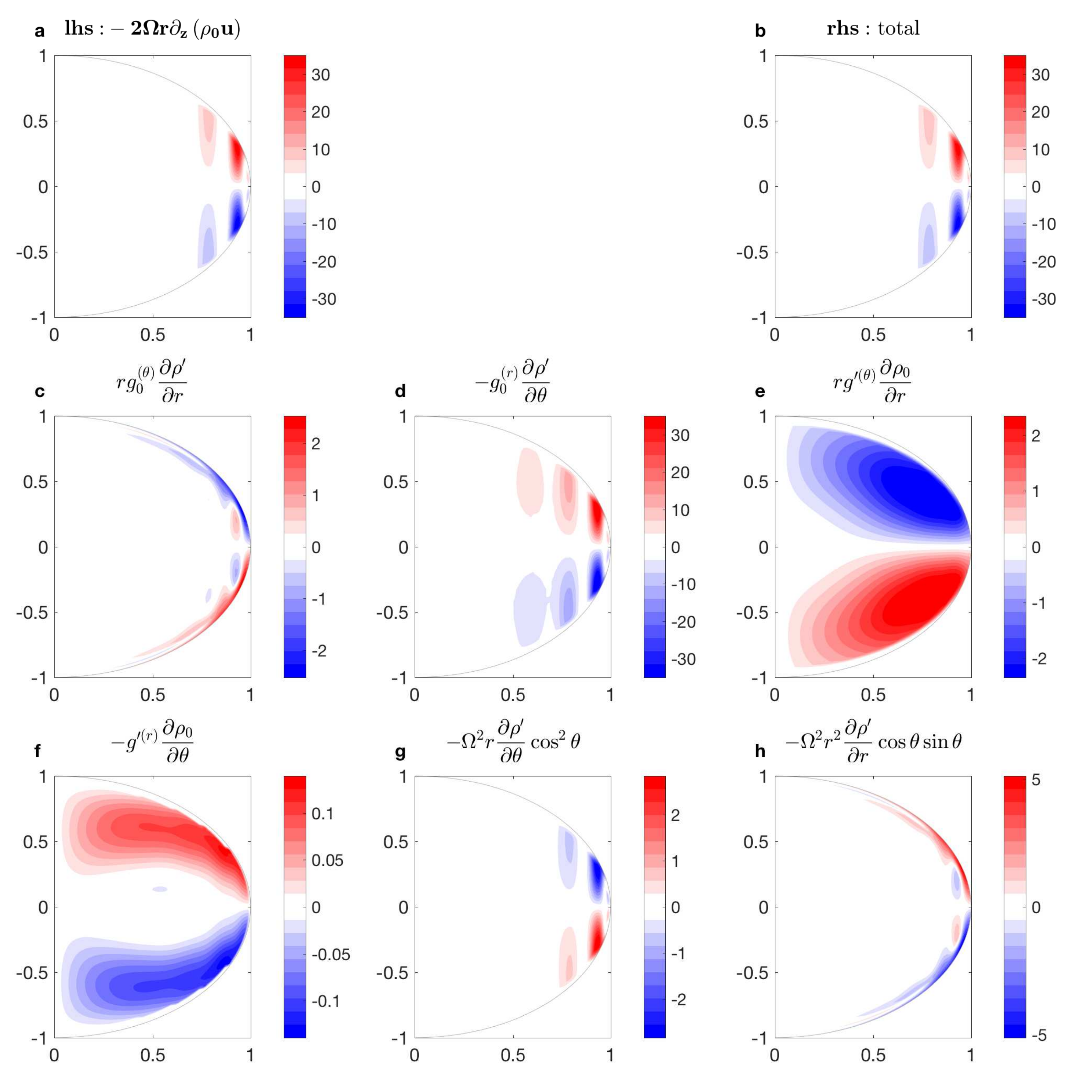}
\par\end{centering}
\caption{\label{fig: full PA - terms}Solution of Eq.~\ref{eq:vorticity equation for dynamic state}
when all terms are kept. (a) $lhs$ term, (b) total $rhs$, (c-h)
the six terms on the $rhs$. Note the different scales in the different
panels.}
\end{figure*}

\begin{figure}
\begin{centering}
\includegraphics[scale=0.38]{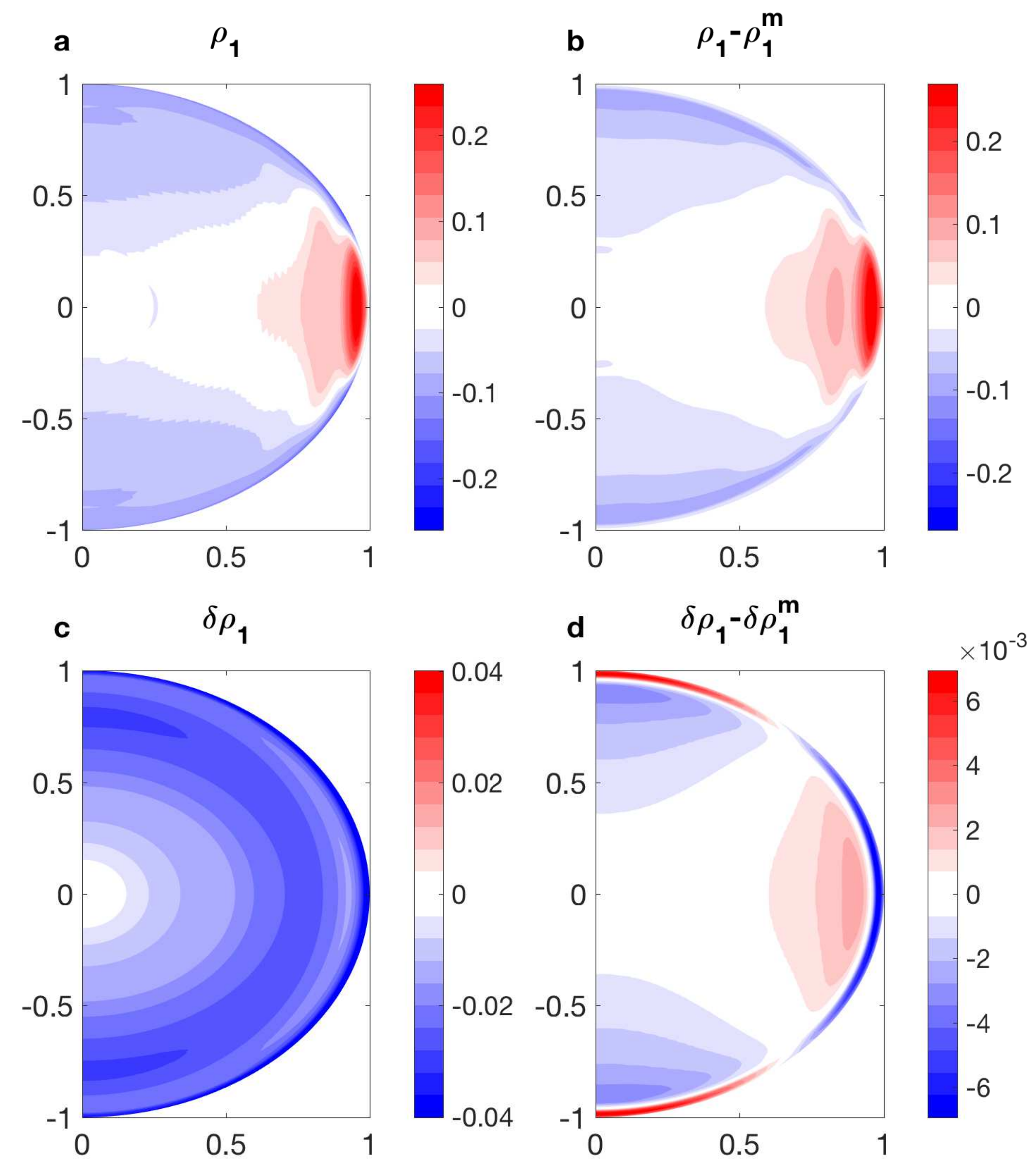}
\par\end{centering}
\caption{\label{fig: drho}Upper panels: the solution for $\rho'$ (a) and
the latitudinal depended part of the solution (b). Lower panels: the
solution for $\delta\hat{\rho}'$ (c) and its latitudinal dependent
part (d). Unlike the eigenvectors (Fig.~\ref{fig: null modes example}),
the solution for the null space is large scale in both radius and
latitude. It is negative everywhere (as a result of the need to compensate
for $\delta\hat{\rho}'$, which is positive everywhere). A large scale
latitude depended structure exists, with the highest values close
to the planet surface. All values are in kg$\,$m$^{-3}$. }
\end{figure}

Next, consider each term in Eq.~(\ref{eq:vorticity equation for dynamic state})
as function of radius and latitude (Fig.~\ref{fig: full PA - terms}),
where panel (a) shows the $lhs$, panel (b) the total of the $rhs$,
and panels (c-h) show the individual contribution from the six different
terms on the $rhs$. The dominant term on the $rhs$ balancing the
$lhs$ is the second term (Fig.~\ref{fig: full PA - terms}d), i.e.,
the thermal wind term, whose magnitude is about ten times larger than
any of the the other terms. Below, in section~\ref{subsec:Analysis-of-solution},
we examine less complete solutions each including only some of the
terms in Eq.~(\ref{eq:vorticity equation for dynamic state}), including
the simplified thermal wind approach of \citet{Kaspi2010a} and the
thermal-gravity wind (TGW) approximation of \citet{Zhang2015}. It
is already clear, though, that the TGW term (Fig.~\ref{fig: full PA - terms}e)
is of the same magnitude as several others, so a self-consistent approximation
can either neglect all terms, but the most dominant one as done in
\citet{Kaspi2010a}, or include all other terms as well as done here
in the full, self-consistent thermal wind (FTW) approach. The perturbation
density solution, $\rho'$, is concentrated near the surface to a
large degree (Fig.~\ref{fig: drho}a,b), and therefore terms that
depend on its vertical derivative are also concentrated near the surface.
Terms that depend on gradients of the zeroth order density are characterized
by a larger scale structure (Fig.~\ref{fig: full PA - terms}e,f). 

Figs.~\ref{fig: drho}c,d show the solution to the null space part
of the density, $\delta\rho'$. It is negative everywhere, in order
to compensate for $\hat{\rho}'$ which is generally positive so that
mass is conserved (section~\ref{subsec:Calculating-drho}). The null
space solution $\delta\rho'$ is smaller than the full solution (Fig.~\ref{fig: drho}a)
by an order of magnitude. Furthermore, the latitude-dependent part
of $\delta\rho'$ (Fig~\ref{fig: drho}d), which is the only part
contributing to the gravity moments, is an order of magnitude smaller
than $\delta\rho'$ itself (Fig~\ref{fig: drho}a). This explains
why the contribution of $\delta\rho'$ to the gravitational moments
(Fig.~\ref{fig: moments - full PA}, dashed line) is at least two
orders of magnitude smaller than that of the non null-space part of
the solution. Overall, this analysis shows that this perturbation
approach gives, to leading order, results that are very close to those
of the CMS. 

\subsection{Analysis of solution and comparison to previous approximations\label{subsec:Analysis-of-solution}}

We now assess the contribution of each term on the $rhs$ of the equation
to the density solution (Fig.~\ref{fig: full PA - terms}), and to
the gravitational moments in particular. Since the equation is linear
with respect to $\rho'$ the analysis can be done by solving the equation
when different terms are excluded. Following is a discussion of the
thermal wind approximation of \citep{Kaspi2010a}, and of the thermal-gravity
wind solution of \citet{Zhang2015}.

\subsubsection{Spherically symmetric thermal wind approximation}

The simplest solution to Eq.~(\ref{eq:vorticity equation for dynamic state}),
the thermal wind (TW) approximation, is obtained when assuming that
the static solution is spherically symmetric \citep{Kaspi2010a},
density and gravity as in Figs.~\ref{fig: background solution}a,c)
satisfying $\rho_{0}=\rho_{0}^{m}$, $g_{0}^{(r)}=g_{0}^{(r)m}$,
and $g_{0}^{(\theta)}=0$, and neglecting the gravity anomaly ${\bf g}'$,
so that
\begin{eqnarray}
\rho & = & \rho_{0}(r)+\rho'(r,\theta),\nonumber \\
{\bf g} & = & g_{0}^{(r)}(r).\label{eq:thermal wind}
\end{eqnarray}
These assumptions reduce Eq.~(\ref{eq:vorticity equation for dynamic state})
to, 
\begin{eqnarray}
2\Omega r\partial_{z}\left(\rho_{0}u\right) & = & g_{0}^{(r)}\frac{\partial\rho'}{\partial\theta},\label{eq:vorticity equation - thermal wind}
\end{eqnarray}
where the centrifugal terms drop under the background sphericity assumptions
(see discussion in section~\ref{sec:Discussion-and-conclusion}).
The solution for the anomalous density can be simply found by integrating
the \textit{rhs} of Eq.~(\ref{eq:vorticity equation - thermal wind})
so that,
\begin{eqnarray}
\rho'(r,\theta) & = & \int^{\theta}\frac{2\Omega r}{g_{0}^{(r)}(r)}\partial_{z}\left(\rho_{0}(r)u(r,\theta')\right)d\theta'+\tilde{\rho}'(r),\label{eq:thermal wind solution}
\end{eqnarray}
where $\tilde{\rho}'$ is an unknown integration coefficient that
does not contribute to the gravitational moments (see Eq.~\ref{eq: dynamical zonal harmonics-constant of integration}).
Note that Eq.~(\ref{eq:vorticity equation - thermal wind}) is not
the standard form of the thermal wind equation, e.g.~\citet{Vallis2006},
since it includes $\rho_{0}$ on the \textit{lhs}, and the \textit{rhs}
is not a purely baroclinic term; nonetheless the two forms are equivalent
(see details in \citealp{Kaspi2016}). Solving Eq.~(\ref{eq:thermal wind solution})
and calculating the gravity moments using Eq.~(\ref{eq: dynamical zonal harmonics})
we can compare the solution to the CMS method (Fig.~\ref{fig: moments - TW-TGW}a).
The thermal wind solution follows the full CMS solution with the ratio
between the calculated moments being $0.91,1.44,1.6,1.53,1.56,0.88$
for $J_{2},J_{4},...,J_{12}$, respectively. Note that in order to
maintain the same framework, the equation was solved numerically using
the same methodology as in the full case. Solving the equation using
the method of \citet{Kaspi2010a}, which is much more efficient numerically,
gives the same results up to numerical roundoff.

A variation on this case \citep{Cao2015}, is to allow the background
density $\rho_{0}$, as well as the gravity in the radial direction,
$g_{0}^{(r)}$, to vary with latitude (Fig.~\ref{fig: background solution}b,d).
The gravity in the latitudinal direction is kept zero. The resulting
equation is the same as Eq.~(\ref{eq:thermal wind solution}), but
with the background fields being a function of both radius and latitude,
\begin{eqnarray}
\rho'(r,\theta) & = & \int^{\theta}\frac{2\Omega r}{g_{0}^{(r)}(r,\theta)}\partial_{z}\left(\rho_{0}(r,\theta)u(r,\theta')\right)d\theta'+\tilde{\rho}'(r).\label{eq:thermal wind solution-oblate}
\end{eqnarray}
The solution to this approximation is very similar to the above simplified
thermal wind balance (indistinguishable from the green line in Fig.~\ref{fig: moments - TW-TGW}),
aside from some differences in the higher gravity moments, especially
$J_{12}$ as was also found by \citet{Cao2015}.

\subsubsection{The thermal-gravity wind approximation}

\begin{figure*}
\begin{centering}
\includegraphics[scale=0.4]{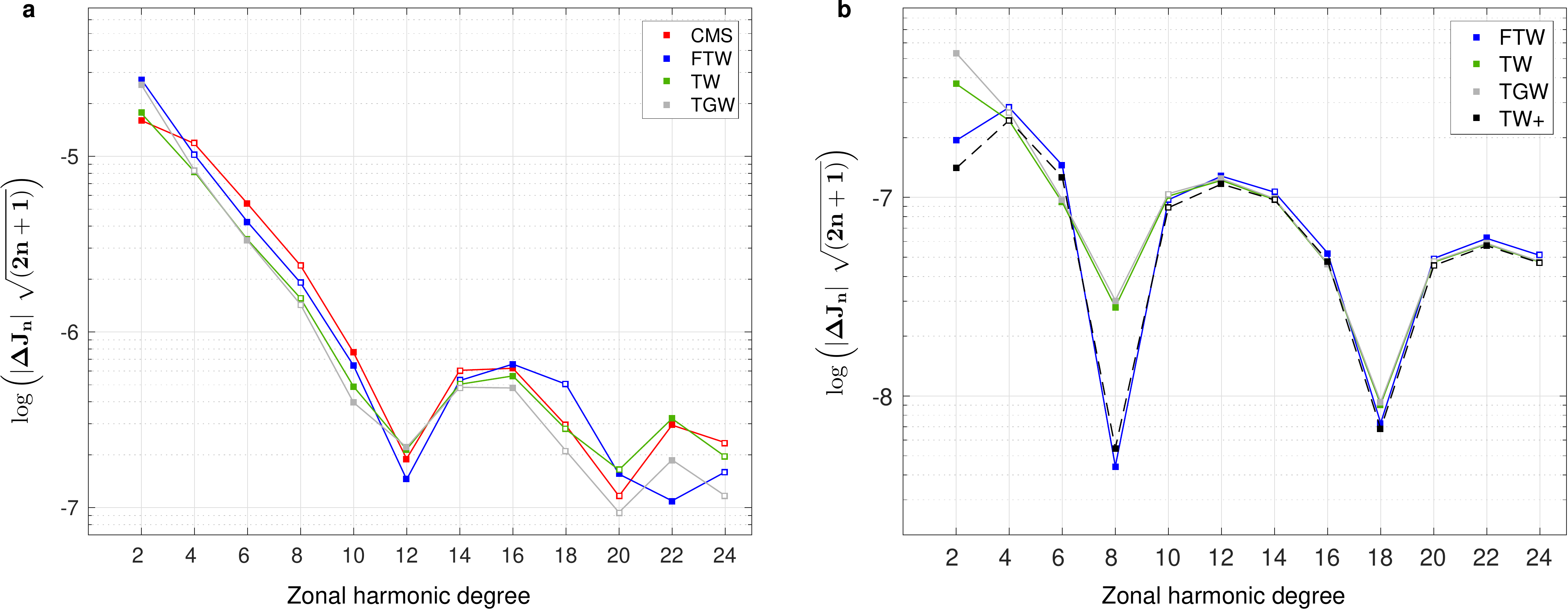}
\par\end{centering}
\caption{\label{fig: moments - TW-TGW}The wind induced gravitational moments
from several limits of Eq.~\ref{eq: dynamical zonal harmonics}.
(a) For barotropic winds, showing the full solution (blue), standard
thermal wind (green), and the thermal-gravity (gray). The CMS solution
is shown in red for comparison. All solutions are quite similar, indicating
that the simple TW approximation produces essentially the correct
solution shown by the fuller approximations. (b) For a baroclinic
case (with a wind-decay scale of 1000km). Shown are full solution
(blue), standard thermal wind (green), and the thermal-gravity (gray)
and a solution where the terms depending on $\partial\rho_{1}/\partial r$
are included (see text for discussion of the TW+ case).}
\end{figure*}

Next, we examine the contribution of the anomalous gravity to the
solution, termed by \citet{Zhang2015} the thermal-gravity equation
(TGW). They suggested that since the density perturbations $\rho'$
result also in perturbations to the gravity field $\mathbf{g}\mathbf{'}$,
these in turn affect the solution, and therefore need to be included
in the balance. As in \citet{Zhang2015} we assume the background
state to vary with radius only, so that,
\begin{eqnarray}
\rho & = & \rho_{0}(r)+\rho'(r,\theta),\label{eq:thermal gravity density}\\
{\bf g} & = & g_{0}^{(r)}(r)+\mathbf{g'}(r,\theta).\label{eq:thermal gravity wind}
\end{eqnarray}
These assumptions reduce Eq.~(\ref{eq:vorticity equation for dynamic state})
to,
\begin{eqnarray}
-2\Omega r\partial_{z}\left(\rho_{0}u\right) & = & -g_{0}^{(r)}\frac{\partial\rho'}{\partial\theta}+r\frac{\partial\rho_{0}}{\partial r}g'{}^{(\theta)}.\label{eq:vorticity equation - thermal gravity}
\end{eqnarray}
This equation cannot be easily integrated in $\theta$, and needs
to be solved numerically \citep{Zhang2015}. The resulting gravity
moments are shown in Fig.~\ref{fig: moments - TW-TGW}a (gray), together
with the thermal wind solution and the full perturbation method solution.
The overall effect of the term added in Eq.~(\ref{eq:vorticity equation - thermal gravity})
relative to the simplified thermal wind is small. It is mostly apparent
in $J_{2}$ which increases by $54\%$. The effect on higher moments,
not calculated in \citet{Zhang2015}, is much smaller. The small effect
of the additional term in TGW approximation is already clear from
the magnitude of the relevant term Fig.~\ref{fig: full PA - terms}e
(repeating Fig.~\ref{fig: full PA - terms} with a radially dependent
background state gives a similar structure and magnitude to that shown
in Fig.~\ref{fig: full PA - terms}a,c,e). Note that solving the
equation with background fields that are both radially and latitudinally
dependent shows similar results in the gravity moments.

\subsection{The perturbation method in the more general case of baroclinic winds\label{subsec:analysis-baroclinic-winds}}

\begin{figure*}
\begin{centering}
\includegraphics[scale=0.5]{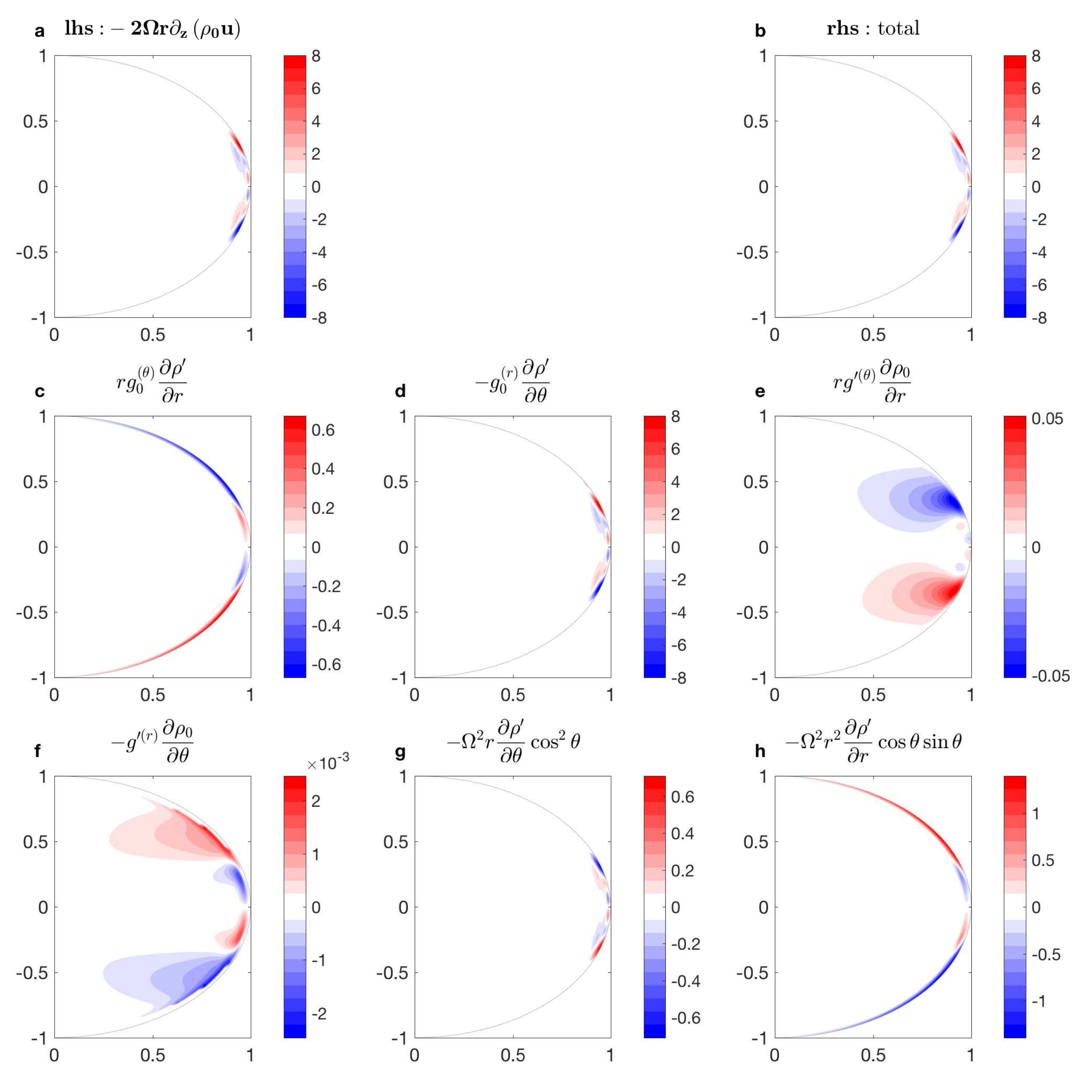}
\par\end{centering}
\caption{\label{fig: full PA - baroclinic}Solution of Eq.~\ref{eq:vorticity equation for dynamic state}
for a baroclinic case (depth of winds is 1000~km) when all terms
are included. (a) $lhs$ term, (b) total $rhs$, (c-h) the six terms
on the $rhs$. Values of $\rho$' are in kg$\,$m$^{-3}$.}
\end{figure*}

So far we have focused only on barotropic cases since the CMS solution,
which we used as our reference, may only be obtained for barotropic
winds. The FTW can be used also to analyze baroclinic winds, which
are considered in this section. Using the baroclinic winds described
in section~\ref{subsec:Prescribed-winds} (decay depth of wind is
1000km), we repeat the above calculations of the density field and
gravity moments. The gravitational moments for this case are shown
in Fig.~\ref{fig: full PA - terms}b, and the individual terms in
the FTW equation are shown in Fig.~\ref{fig: full PA - baroclinic}.

The solution for the gravitational moments shows that the TW \citep{Kaspi2010a}
and TGW \citep{Zhang2015} solutions are again remarkably similar,
apart from $J_{2}$. The solutions for both of these approximations
are similar to the fuller FTW approximation, except for the moments
$J_{2}$, $J_{6}$ and $J_{8}$. In particular, the fuller solution
for $J_{8}$ is an order of magnitude smaller than both cruder approximations,
underlining the importance of considering the additional physical
effects included in this paper. In this case, there is a significant
difference between TW (green) and the similar TGW (gray) on the one
hand, and FTW (blue) on the other. The main reason for this difference
are the two terms involving $\partial\rho'/\partial r$. This is shown
by the dash black curve denoted TW+, where we have used the TW solution
plus the terms shown in panels c and h in Fig.~\ref{fig: full PA - baroclinic},
which both involve the radial derivative of the perturbation density.
The importance of these terms is a direct consequence of the structure
of the perturbation density solution, which tends to be strongly concentrated
near the upper surface. This surface enhancement is not surprising
given that the wind forcing decays rapidly away from the surface in
this baroclinic case. This also implies that all terms in the equation
tend to be more concentrated near the surface than in the barotropic
case (compare Fig.~\ref{fig: full PA - terms} and Fig.~\ref{fig: full PA - baroclinic}).
For this baroclinic case, the TGW term (panel e) is negligible relative
to most other terms considered here. Note also, that the solution
of the null space (section~\ref{subsec:Calculating-drho}) relies
on the barotropically based polyrtopic equation, therefore some inconsistencies
might arise due to that. This, however, should not affect much the
solutions since the null space contribution to the solution is small.

\section{Discussion and conclusion\label{sec:Discussion-and-conclusion}}

In the traditional approximation for terrestrial planets the centrifugal
term is often merged with gravity in the momentum equation, by choosing
the vertical direction to be that perpendicular to the planet's geopotential
surface and defining an effective gravity. This is then traditionally
followed by approximating the planet as a sphere, so that the vertical
direction coincides with the radial direction, and thus effectively
neglecting the horizontal component of the centrifugal term. \textcolor{black}{It
is important to note that this centrifugal term is not smaller than
the Coriolis term even for the Earth case, but because of the nearly
spherical shape of Earth, this approximation allows trading a large
dynamical component in the momentum balance with a small geometric
error }\citep[section 2.2]{Vallis2006}. This approximation has proven
to hold well for Earth and other terrestrial planets. On the giant
planets, the oblateness is not small (6.5\% and 9.8\% on Jupiter and
Saturn, respectively, compared to 0.3\% on Earth). Therefore, the
contribution of the centrifugal (fifth and sixth terms on the \textit{rhs}
of Eq.~\ref{eq:vorticity equation for dynamic state}) and self-gravitation
terms (third and fourth terms on the \textit{rhs} of Eq.~\ref{eq:vorticity equation for dynamic state})
can potentially lead to significant contributions to the momentum
balance, and therefore may alter thermal-wind balance as well. The
goal of this study is to assess the importance of these terms in a
fluid planet.

By solving numerically the full second order momentum equation in
our perturbation approach, which includes the original thermal-wind
balance terms (\textit{lhs} term and second term on \textit{rhs} in
Eq.~\ref{eq:vorticity equation for dynamic state}), self-gravity
terms, centrifugal terms, and other non-spherical contributions (first
term on \textit{rhs} of Eq.~\ref{eq:vorticity equation for dynamic state}),
we have shown that the original thermal wind balance is still the
leading order. In the barotropic limit, the thermal wind results,
with the various higher order contributions, are systematically compared
to results from the oblate concentric Maclaurin spheroids (CMS) model.
In the context of recent studies that argue that additional terms
are important in the balance for calculating the gravitational moments
\citep{Zhang2015,Cao2015}, we show that to leading order these terms
are negligible, and have a small contribution to the gravity moments.
Consistently with \citet{Zhang2015}, we find that the self-gravity
term (TGW) increases the value of $J_{2}$, though it does not bring
the thermal wind $J_{2}$ closer to the CMS result. This terms has
a negligible contribution to all higher harmonics, which were not
discussed in \citet{Zhang2015}.

We conclude therefore that while more complete solutions are possible,
as we do in this study, the traditional thermal-wind gives a very
good approximation to the balance between the wind shear and the density
gradients, and integrating it gives a very good approximation to the
gravity moments. Particularly, taking into account the accuracy of
the Juno and Cassini measurements, this gives an excellent approximation.
Quantitatively, for the barotropic cases, its results differ by at
most a factor of 1.6 compared to the full solution. This difference
is small considering the other uncertainties of the interior flow.
For the baroclinic case, where wind structures decay rapidly near
the surface, terms involving the radial derivative of the perturbation
density become more important for calculating the gravity moments.

The main advantage of using the thermal wind (TW) model compared to
full, self-consistent thermal wind (FTW) is numerical. While the TW
equation (\ref{eq:thermal wind solution}) allows for local calculation
of the density from the wind, the FTW equation (\ref{eq:vorticity equation for dynamic state})
is an integro-differential equation that needs to be solved globally.
It is mostly complicated from the need to integrate the dynamical
density ($\rho'$ ) globally to calculate the dynamical self-gravity
($\mathbf{g'}$). The TW approximation allows therefore using much
higher resolution, which is necessary for resolving the high order
moments. As a consequence of the simplicity of the TW model, more
sophisticated and numerically demanding methods can be applied in
order to find the best matching wind field \textit{given} the gravity
measurements \citep{Galanti2016,Galanti2016c}. The inevitability
of the solution using the TW model is a major advantage for the upcoming
analysis of the Juno and Cassini data. Given the extremely small contribution
of the null space to the overall solution, we expect that the more
complete FTW model would also be invertible, still the computational
challenge involved is much greater.

In summary, deciphering the effect of the atmospheric and internal
flows from the measured gravity spectrum of Jupiter and Saturn provides
a major challenge. The methods suggested to date have been either
limited to barotropic cases \citep[e.g.,][]{Hubbard1982,Hubbard1999,Hubbard2012,Kong2012,Hubbard2014},
or approximations limited to spherical symmetry or partial solutions
\citep[e.g.,][]{Kaspi2010a,Zhang2015,Cao2015}. Here, we have developed
a self-consistent perturbation approach to the thermal wind balance
that incorporates all physical effects, including the effects of oblateness
on the dynamics and the gravity perturbation induced by the flow itself.
The full self-consistent perturbation approach to the thermal wind
balance considered here allows to objectively examine the role of
different physical processes, allows obtaining and even more accurate
approximation by proceeding to higher order perturbation corrections
(Appendix~\ref{subsec:Appendix integration-constant}), and allows
interpreting the expected Juno and Cassini observations in a more
complete way than was possible in previous approaches, thus maximizing
the benefits of these observations.

~

\textit{Acknowledgements:} We thank the Juno science team interiors
working group for valuable discussions. ET is funded by the NSF Physical
Oceanography program, grant OCE-1535800, and thanks the Weizmann Institute
of Science (WIS) for its hospitality during parts of this work. YK
and EG acknowledge support from the Israeli Ministry of Science (grant
45-851-641), the Minerva foundation with funding from the Federal
German Ministry of Education and Research, and from the WIS Helen
Kimmel Center for Planetary Science.

\bibliographystyle{apalike}
\bibliography{yohaisbib}

\appendix

\section{Appendices}

\subsection{Higher order perturbation equations\label{subsec:Appendix Higher-order-perturbation}}

We write here the perturbation equations to the second order to demonstrate
how the results of our approach can be made more accurate if needed.
The momentum equation (\ref{eq:momentum-1}) is, 

\begin{eqnarray}
2\boldsymbol{\Omega}\mathbf{\times}\left(\rho\mathbf{u}\right) & = & -\nabla p-\rho{\bf g}-\rho\boldsymbol{\Omega}\times\boldsymbol{\Omega}\mathbf{\times r}.
\end{eqnarray}
Writing the density as $\rho=\rho_{0}+\rho_{1}+\rho_{2}$, and substituting
into the above equation, treating $\rho_{1}$ as an order $\epsilon$
correction and $\rho_{2}$ as an order $\epsilon^{2}$ correction,
we can separate the different orders to find equations for each correction
order. Note that in the previous sectinos we denote $\rho_{1}$ as
$\rho'$. We view, as defined earlier, the zeroth order balance as
the balance without the effects of the winds, so that our zeroth order
equation is,

\begin{eqnarray}
0 & = & -\nabla p_{0}-\rho_{o}g_{0}-\rho_{0}\boldsymbol{\Omega}\times\boldsymbol{\Omega}\mathbf{\times r}.
\end{eqnarray}
The winds enter at the first order, where the equation is

\begin{eqnarray}
2\boldsymbol{\Omega}\mathbf{\times}\left(\rho_{0}\mathbf{u}\right) & = & -\nabla p_{1}-\rho_{o}{\bf g}_{1}-\rho_{1}{\bf g}_{0}-\rho_{1}\boldsymbol{\Omega}\times\boldsymbol{\Omega}\mathbf{\times r},
\end{eqnarray}
and the second order correction is then obtained by solving
\begin{eqnarray}
2\boldsymbol{\Omega}\mathbf{\times}\left(\rho_{1}{\bf u}\right) & +\text{\ensuremath{\rho_{1}{\bf g}_{1}}}= & -\nabla p_{2}-\rho_{o}{\bf g}_{2}-\rho_{2}{\bf g}_{0}-\rho_{2}\boldsymbol{\Omega}\times\boldsymbol{\Omega}\mathbf{\times r}.
\end{eqnarray}
In the second order perturbation equation we moved all terms that
are known from previous orders to the \textit{lhs}. This equation
is again solved by taking a curl and then taking care of the integration
constant (null space solution) if needed. Because the second order
equation is generally similar to the first order equation, its numerical
solution follows the same approach and should not pose significant
additional difficulties. While this procedure should be formally done
in a non-dimensional form, we present here the dimensional equations
for clarity. The small parameter in this expansion when it is done
in non-dimensional form is the Rossby number $Ro=(\Omega U)/L$, where
$U$ is a scale for the velocity and $L$ the horizontal length scale
of the relevant motions.

\subsection{Integration constant and null space\label{subsec:Appendix integration-constant}}

The objective of this appendix is to show how the integration constant
encountered in the thermal wind approach \citep[e.g., ][]{Kaspi2010a,Zhang2015}
can be calculated. This is meant to aid the understanding of our approach
to solving for the null space of the more general solution considered
in the main text. The main message of this appendix is that the integration
constant may be determined by adding the missing physics of the polytropic
equation, and by requiring the total mass of the density perturbation
to vanish. The momentum equation for the simple example is,

\begin{equation}
0=-\nabla p-\mathbf{g}\rho+\Omega^{2}\rho r\cos\theta\hat{r}_{\perp}.
\end{equation}
We treat the effects of rotation as a perturbation and the leading
order balance is then hydrostatic, with the corresponding fields being
a function of radius alone ($\rho_{0}=\rho_{0}\left(r\right)$), 
\begin{equation}
0=-\nabla p_{0}-\mathbf{g}_{0}\rho_{0}.
\end{equation}
The next order contains the deviations in density and gravity due
to rotation, 
\begin{equation}
0=-\nabla p_{1}-\mathbf{g}_{1}\rho_{0}-\mathbf{g}_{0}\rho_{1}+\Omega^{2}\rho_{0}r\cos\theta\hat{r}_{\perp}.\label{eq:appndix-O(eps)-momentum}
\end{equation}
Taking the curl, 
\begin{eqnarray}
0 & = & -g_{1}^{(\theta)}r\frac{\partial\rho_{0}}{\partial r}+g_{0}\frac{\partial\rho_{1}}{\partial\theta}-\Omega^{2}r^{2}\frac{\partial\rho_{0}}{\partial r}\cos\theta\sin\theta,\label{eq:vorticity_zonal_1st_order_2}
\end{eqnarray}
substituting the expression for the gravity fields, and integrating
over $\theta$,

\begin{eqnarray}
-\frac{1}{2}\Omega^{2}r^{2}\frac{\partial\rho_{0}}{\partial r}\cos^{2}\theta & = & g_{0}\rho_{1}+2\pi G\frac{\partial\rho_{0}}{\partial r}\iint\frac{\mathbf{\rho_{_{1}}}(r',\theta')r\mathbf{'^{2}}\sin(\theta')}{\langle|\mathbf{r-}\mathbf{r}'|\rangle}dr'd\theta'+C(r),\label{eq:appendix-equation}
\end{eqnarray}
where $C(r)$ is the unknown integration constant to be solved for,
and the \textit{lhs} is a known forcing term due to rotation by to
the already known zeroth order solution. The perturbation density
can be expressed as the sum of $\hat{\rho}_{1}(r,\theta)$ that solves
the above equation with the integration constant set to zero, and
$\delta\rho_{1}(r)$ that satisfies the same equation with a zero
on the \textit{lhs} and with the integration constant. Together, $\rho_{1}=\hat{\rho}_{1}(r,\theta)+\delta\rho_{1}(r)$
solves Eq.~\ref{eq:appendix-equation}.

Given that we took the curl of the momentum equation, $\hat{\rho}_{1}$
which solves the above equation without $C(r)$ satisfies the original
momentum equation up to a gradient of some function which we may write
as $\hat{p}_{1}+\delta p_{1}$, 
\[
0=-\nabla(p_{1}+\delta p_{1})+\rho_{0}\mathbf{\hat{g}}_{1}+\hat{\rho}_{1}\mathbf{g}_{0}+\rho_{0}\Omega^{2}r\cos\theta\hat{r}_{\perp}.
\]
Next, take the difference between the momentum equation for $\rho_{1}$
and for $\hat{\rho}_{1}$, we find,
\begin{equation}
0=-\nabla\delta p_{1}+\rho_{0}\mathbf{\delta g}_{1}+\delta\rho_{1}\mathbf{g}_{0}.\label{eq:O(eps)-correction}
\end{equation}
The second and third terms in \ref{eq:O(eps)-correction} are functions
of radius $r$ only. Therefore we expect the pressure term $-\nabla\delta p_{1}$
to also be a function of the radius only. Furthermore, assuming now
that the polytropic relation holds for the perturbation pressure and
density, we have
\[
\delta p_{1}=Kn\rho_{0}\delta\rho_{1}
\]
which leads to an equation for the unknown perturbation density,
\[
0=-\nabla\left(Kn\rho_{0}\delta\rho_{1}\right)+\rho_{0}\mathbf{\delta g}_{1}+\delta\rho_{1}\mathbf{g}_{0}.
\]
Substituting the expression for the gravity, 
\begin{eqnarray}
0 & = & -\frac{\partial}{\partial r}\left(Kn\rho_{0}\delta\rho_{1}\right)+\rho_{0}2\pi G\frac{\partial}{\partial r}\iint\left\langle \frac{1}{|\mathbf{r-}\mathbf{r}'|}\right\rangle \delta\rho_{1}(r')r\mathbf{'^{2}}\sin(\theta')dr'd\theta'+\delta\rho_{1}\mathbf{g}_{0},\label{eq:order1-integration-correction}
\end{eqnarray}
and substituting the zeroth order solutions for $\rho_{0},$ $p_{0}$
and $g_{0}$ (\citet{Zhang2015}, see notation there), and after some
more rearrangement and integration in $\xi$ we get, 
\begin{eqnarray}
0 & = & -\delta\rho_{1}+\frac{1}{2}\iint\left\langle \frac{1}{|\mathbf{\xi-}\xi'|}\right\rangle \delta\mathbf{\rho_{_{1}}}(\xi')\xi\mathbf{'^{2}}\sin(\theta)d\xi'd\theta+D\label{eq:order1-integration-correction-2-1-1}
\end{eqnarray}
where $D$ is the constant of integration. Use the expansion for the
average over longitude \citep{Zhang2015}, and the fact that the integral
over all Legendre polynomials but the first vanish,
\begin{eqnarray}
0 & = & -\delta\rho_{1}(\xi)+\int_{0}^{\pi}f_{0}(\xi,\xi')\delta\mathbf{\rho_{_{1}}}(\xi')\xi\mathbf{'^{2}}d\xi'+D,\label{eq:delta-rho1-after-theta-avg}
\end{eqnarray}
where,
\[
f_{0}(\xi,\xi')=\begin{cases}
\frac{1}{\xi} & \xi'\le\xi\\
\frac{1}{\xi'} & \xi'>\xi
\end{cases}.
\]
A solution may be found by assuming a Frobenius-form solution,
\[
\delta\rho_{1}(\xi)=\sum_{m=0}^{\infty}a_{m}\xi{}^{m+r}.
\]
Now find what is $a_{m}.$ Write Eq.~(\ref{eq:delta-rho1-after-theta-avg})
explicitly
\begin{eqnarray}
0 & = & -\sum_{m=0}^{\infty}a_{m}\xi^{m+r}+\sum_{m=0}^{\infty}\int_{0}^{\xi}\frac{1}{\xi}a_{m}\xi'^{m+r}\xi\mathbf{'^{2}}d\xi'+\sum_{m=0}^{\infty}\int_{\xi}^{\pi}\frac{1}{\xi'}a_{m}\xi'^{m+r}\xi\mathbf{'^{2}}d\xi'+D,\label{eq:delta-rho1-after-theta-avg-explicit-2}
\end{eqnarray}
and performing the integral and collecting powers, while also defining
$a_{-1}=a_{-2}=0$ ,
\begin{align}
0 & =\sum_{m=0}^{\infty}\left(-a_{m}+\frac{1}{\left(m+1+r\right)}a_{m-2}-\frac{1}{m+r}a_{m-2}\right)\xi^{m+r}+\sum_{m=0}^{\infty}\frac{1}{m+2+r}a_{m}\pi^{m+2+r}+D.\label{eq:power-series-equation}
\end{align}
The coefficient of $\xi^{m+r}$ should vanish for all $m$, giving
us the recursion relation. Multiplying by $(m+r)(m+r+1)$ and considering
the $m=0$ and $m=1$ cases, and we find that there are two possible
solutions for the Frobenius power, $r=0$ or $r=-1$. The first leads
to the recursion relation,
\begin{eqnarray*}
a_{m} & = & -\frac{1}{m(m+1)}a_{m-2}.
\end{eqnarray*}
We term this solution $F_{1}(\xi).$ The second solution, with $r=-1$,
leads to
\begin{eqnarray}
a_{m} & = & -\frac{1}{(m-1)m}a_{m-2},\label{eq:delta-rho1-after-theta-avg-explicit-1-2-1-1-1}
\end{eqnarray}
which is the recursion relation for cosine. Therefore,

\[
\delta\rho_{1}=C_{1}F_{1}+C_{2}\frac{\cos(\xi)}{\xi}.
\]
The $\cos(\xi)/\xi$ solution is not physical because it diverges
at $\xi=0$ and we conclude that $C_{2}=0$. Considering the coefficients
of $\xi^{0}$ in Eq.~\ref{eq:power-series-equation} we find $a_{0}$
in terms of the unknown integration constant $D$. We may then use
$D$ to satisfy the constraint that volume integral over the total
perturbation $\rho_{1}(\xi,\theta)+\delta\rho_{1}(\xi)$ should vanish,
as done in the manuscript for the fuller problem considered there.
\end{document}